\documentclass[12pt]{iopart}
\usepackage{epsfig}  
\begin{document}

\title[Kaon-nucleon and D-nucleon scattering in the quark model]{Kaon-nucleon and D-nucleon scattering in the quark model, including spin-orbit interactions}

\author{Noel Black \footnote[3]{nblack@utk.edu}}

\address{Department of Physics and Astronomy \\ 
University of Tennessee, Knoxville, TN, 37996, USA}

\begin{abstract}

Interactions of charmed and strange mesons with baryonic matter can be calculated in the nonrelativistic 
quark potential model. 
For $KN$ scattering data exists, and the theoretical results for S-waves are in approximate agreement
with experiment. Here we apply the same model to the scattering of open-charm ($D$) mesons by
nucleons, and give quark model predictions for $DN$ scattering amplitudes. Spin-orbit forces in
$KN$ and $DN$ will also be discussed. 

\end{abstract}




\section{Introduction}

$KN$ scattering is ideally suited for studying the origins of the nonresonant ``nuclear" force.
Conventional s-channel baryon resonance production is excluded because there is no valence annihilation,
and one-pion exchange is forbidden because there is no three-pseudoscalar vetex.  We may therefore
study the nonresonant part of hadron scattering in relative isolation, uncomplicated by one-pion
exchange.
In this work we calculate $KN$ phase shifts at Born order in a quark exhange model with one gluon
exchange (OGE) and linear scalar confinement. 
The dominant hyperfine term has given very good results in S-wave for I=2 $\pi \pi$ [1], 
I=3/2 $K \pi$ [2],
I=1 $KK$[1],
$KN$ [3],
and short-ranged $NN$ [4] scattering.  
Here we include the spin-orbit and subdominant spin-independent contributions, and give results for
higher-L waves.
$KN$ scattering is an excellent place to test our model since data exists and there is a
large spin-orbit effect, especially evident in the P-waves. 
Understanding the spin-orbit effect here will be important for applications to other hadronic interactions,
such as $DN$. 
$KN$ scattering is almost entirely elastic below $K \Delta$ threshold,
and the Born approximation is expected to be relevant since the interaction is known experimentally
to be relatively weak.

\section{Experiment}

There are some basic experimental features of $KN$ scattering which are generally agreed upon.  
The I=1 channel is relatively well determined from $K^+ P$ scattering.  
Using the conventional $L_{I \; 2J}$ notation, the $S_{11}$ and $P_{11}$
channels are repulsive and the $P_{13}$ channel is attractive.  I=0 is more in doubt because of inherent
difficulties in the experimental analysis.  
The I=0 $KN$ scattering amplitudes have been extracted from $K\,d$
scattering, using models of the deuteron breakup process and form factors.
It is clear, however, that the $P_{01}$ channel is strongly 
attractive (in fact anomalously so).
Inelasticities are known to be large above inelastic thresholds, in particular in the $P_{01}$ wave.  
For reference, $K \Delta$ (I=1 $KN$ only) opens at
$P_{lab}=0.86$~GeV, $K^* N$ opens at $P_{lab}=1.08$~GeV and $K^* \Delta$ opens 
at a much higher $P_{lab}=1.74$~GeV.

\section{Calculation}

The fundamental interactions in QCD are between quark and gluon constituents.  
The quark-quark OGE $T_{fi}$ is derived as the nonrelativistic reduction to order $P^2/m^2$ of the Feynman
amplitude for two quarks to exchange a gluon with phenomenological strength $\alpha_s$.  The result is 
the usual Breit-Fermi Hamiltonian, which is well known from atomic physics, and includes the
dominant spin-spin hyperfine interaction as well as Coulomb, spin-orbit, tensor, and subdominant
spin-independent contributions.
The confining interaction is modelled as a linear scalar interaction with strength b
(the string tension)
between a pair of quarks, and yields linear and spin-orbit contributions, as well as smaller
spin-independent terms.
The quark-quark interaction is followed by quark interchange, so that the outgoing hadrons emerge as color 
singlets [5].

The scattering arises from the interactions between quarks in different hadrons,
and the interaction Hamiltonian $H_{int}$ is the sum of quark-quark interactions between all such
pairs of quarks.
To Born order the scattering amplitude is given by the matrix element of $H_{int}$
between external hadron states, which ideally should be eigenstates of the respective free Hamiltonians.
We approximate the true wavefunctions by Gaussian forms for calculational 
simplicity, as these lead to analytical results for the scattering amplitudes.  
This approximation has been tested in previous calculations [6].
The parameters are relatively well determined from quark-model phenomenology; we use a reasonably 
standard set (below) for our numerical evaluation.
\begin{equation}
\begin{array}{rcll}
\alpha_s & = &  0.6 & \mbox{strong coupling constant} \nonumber \\
b & = &  0.18 \; \mbox{GeV}^2 & \mbox{string tension} \nonumber \\
\beta & = &  0.4 \; \mbox{GeV} & \mbox{meson wavefunction length scale} \nonumber \\
\alpha & = &  0.4 \; \mbox{GeV} & \mbox{baryon wavefunction length scale} \nonumber \\
m_{u,d} & = &  0.330 \; \mbox{GeV} & \mbox{nonstrange constituent quark mass} \nonumber \\
m_s & = &  0.550 \; \mbox{GeV} & \mbox{strange constituent quark mass}
\end{array}
\end{equation}

\noindent
The phase shifts are calculated from the scattering amplitudes and phase space factors in the
usual way [6].

\section{Results}

Hashimoto [7] has performed the most recent comprehensive single-energy phase shift analysis
we are aware of, and we take his data for comparison.  We also compare our results with those of 
the recent resonating group method (RGM) calculation of Lemaire, \etal [8].  
In the I=1 channel, our results are in reasonable agreement with the RGM calculation.  
Both calculations agree at least qualitatively with the data except in the $P_{13}$ wave, for which they
both give the wrong sign.
In I=0, the absolute values of the calculated phase shifts are in general too small, especially
in the remarkably large $P_{01}$ wave.  
The OGE spin-orbit contributions to our quark model $KN$ phase shifts are shown in Figure 3.  The large 
$P_{01}$ amplitude evident there is perhaps an indication that we have included at least some of the correct
physics.  It is, however, not large enough by itself to account for the data, and is partially cancelled
by the confinement spin-orbit contribution and the negative spin-spin hyperfine term.  

Mukhopadhyay and Pirner [9] have found that the phase shifts calculated from the symmetric part of the 
OGE spin-orbit interaction agree reasonably well with the experimental phase shifts 
at low energies except in the $P_{01}$ wave, which they seriously underestimate. 
The antisymmetric part of the OGE spin-orbit interaction is omitted in their calculation because it gives 
rise to large splittings in the baryon spectroscopy which are not observed.
They similarly conclude that the confinement spin-orbit interaction partially cancels the OGE spin-orbit 
contribution, and the
inclusion of the spin-spin interaction actually
makes the $P_{13}$ phase shift theoretically repulsive.

The opening of inelastic channels is known to be a very large effect in the $P_{01}$ wave, and may be
responsible for the large experimental phase shift.
Coupling to inelastic channels will be treated in a future calculation.  

\section{$DN$ scattering}

Knowledge of vacuum $DN$ scattering amplitudes and observables is important for understanding
open-charm hadronic interactions in hot, dense media, as occur in heavy ion collisions.
We applied our $KN$ 
scattering model to obtain $DN$ elastic phase shifts, which are shown in Figure 4. 
(The only new parameter was the charmed quark mass, taken to be $m_c=1.550$ GeV.)  
Inelasticities may well be large in $DN$, as in $KN$ scattering.
In I=1, the linear and spin-spin hyperfine terms are roughly comparable, and dominate scattering
in the waves shown.
The spin-orbit splitting is small in comparison.  
It is noteable that in I=0 the spin-independent terms (Coulomb and linear) are identically zero and
the spin-orbit interactions dominate the scattering.

\begin{figure}
\epsfig{figure=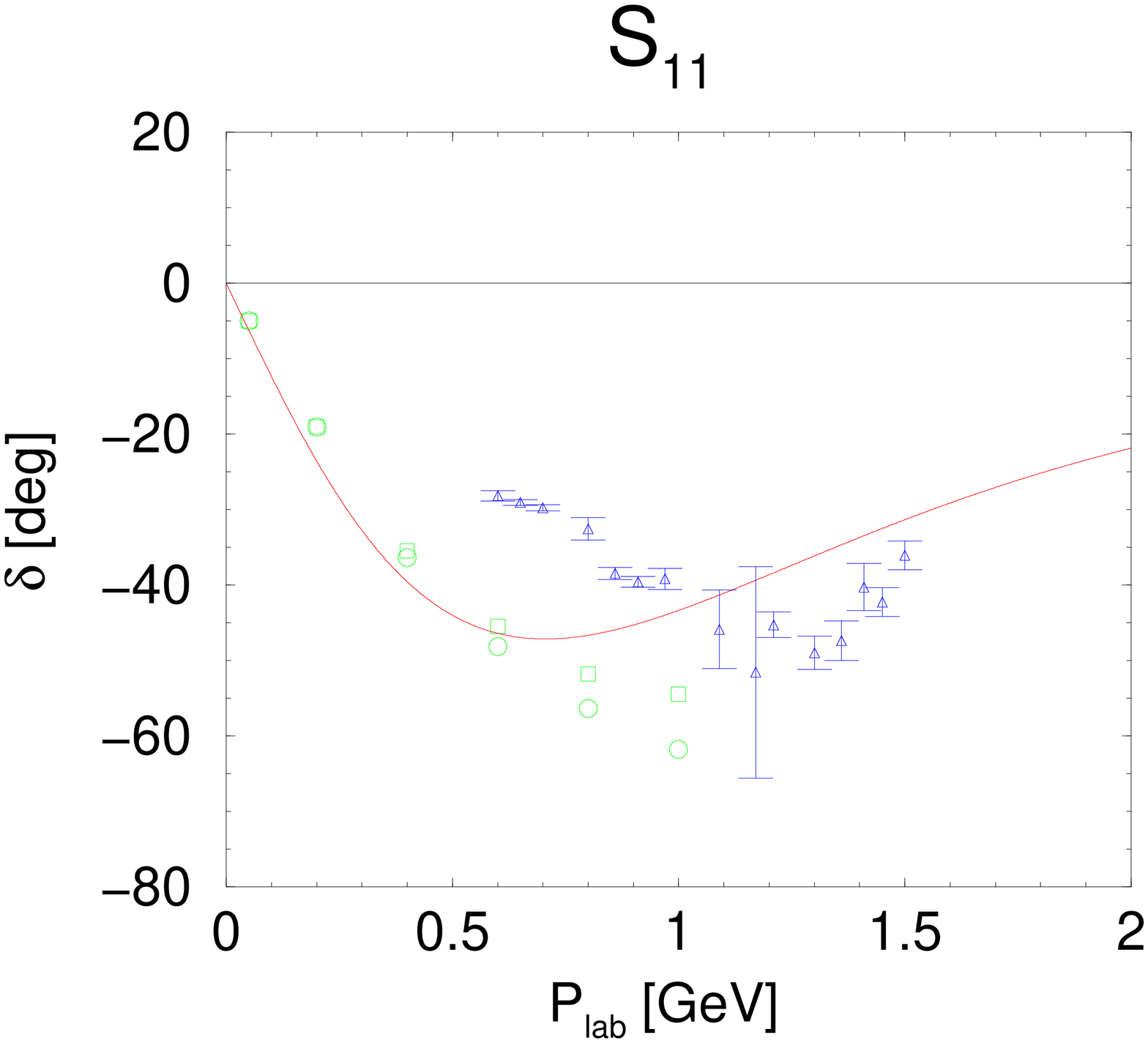,width=1.5in,angle=-90}
\hspace{1cm}
\epsfig{figure=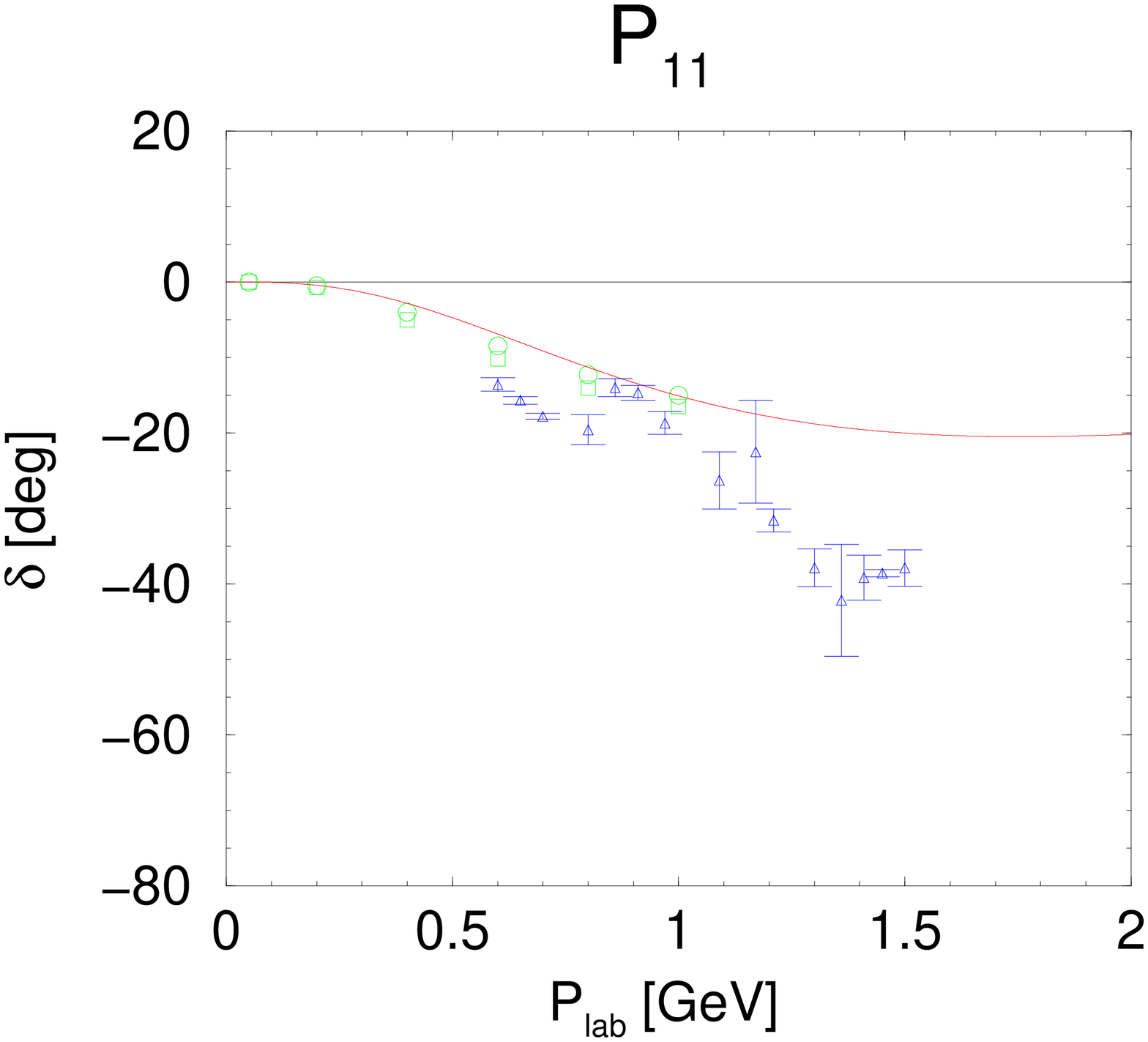,width=1.5in,angle=-90}
\hspace{1cm}
\epsfig{figure=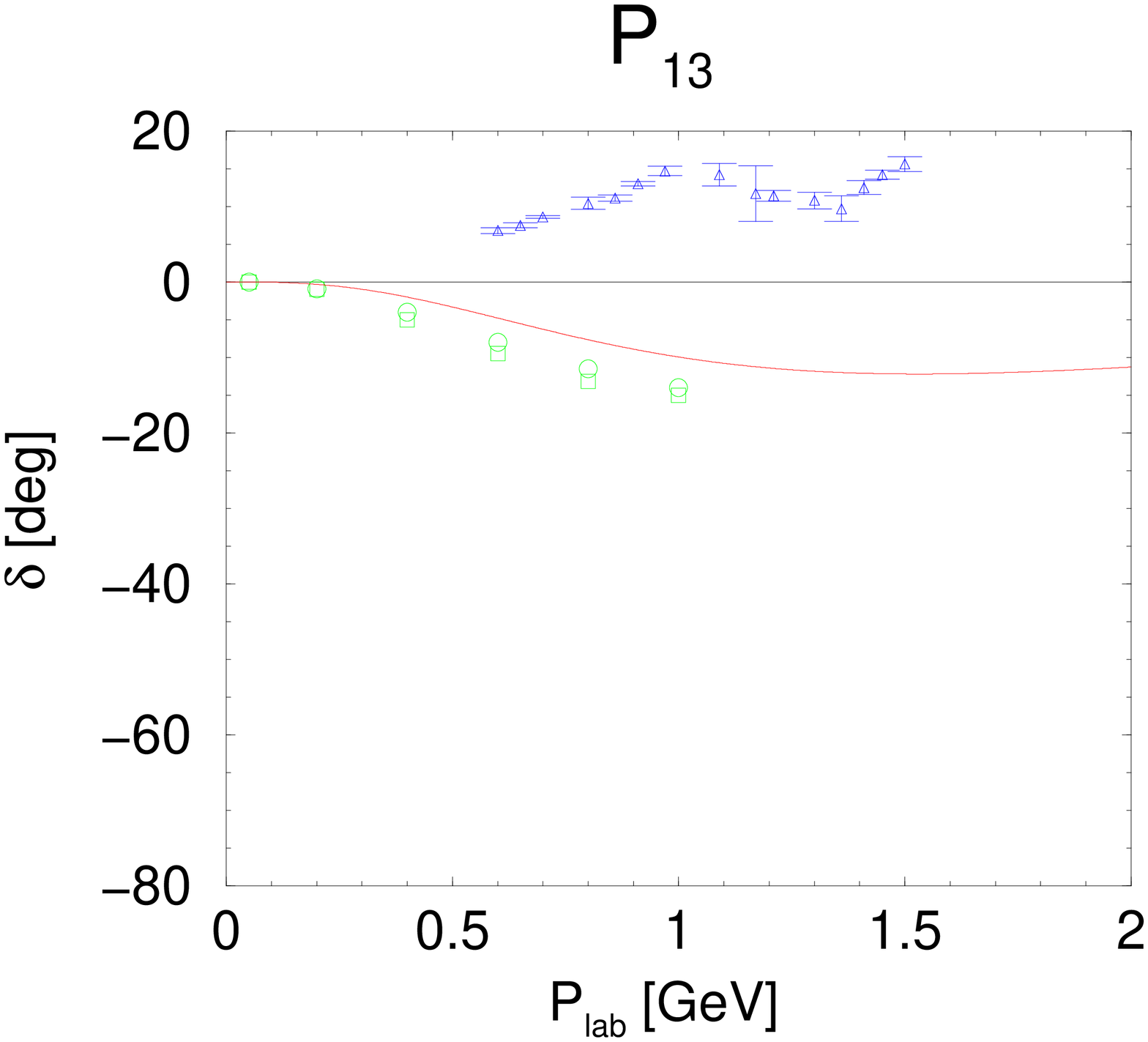,width=1.5in,angle=-90}

\vspace{0.5cm}

\epsfig{figure=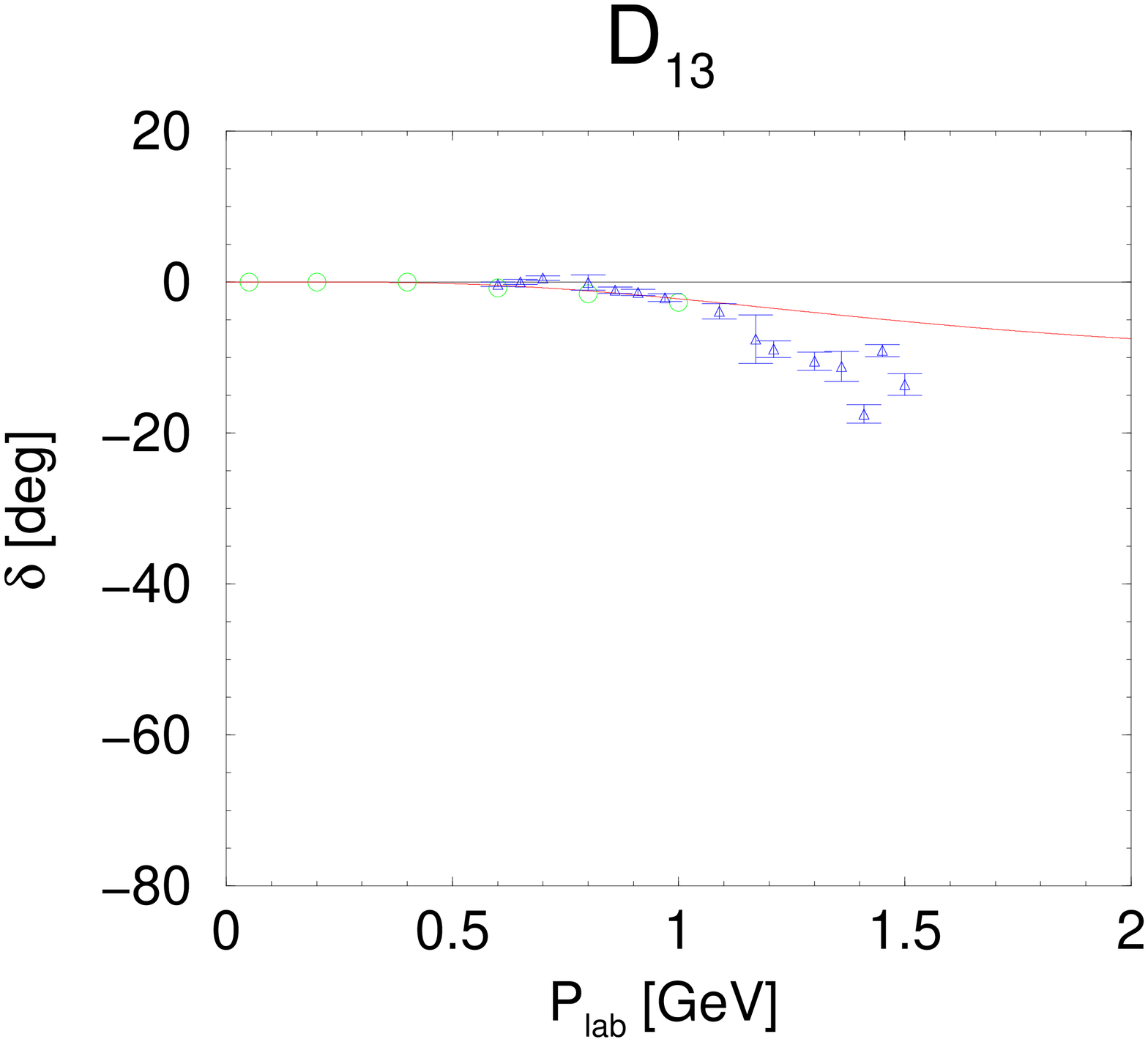,width=1.5in,angle=-90}
\hspace{1cm}
\epsfig{figure=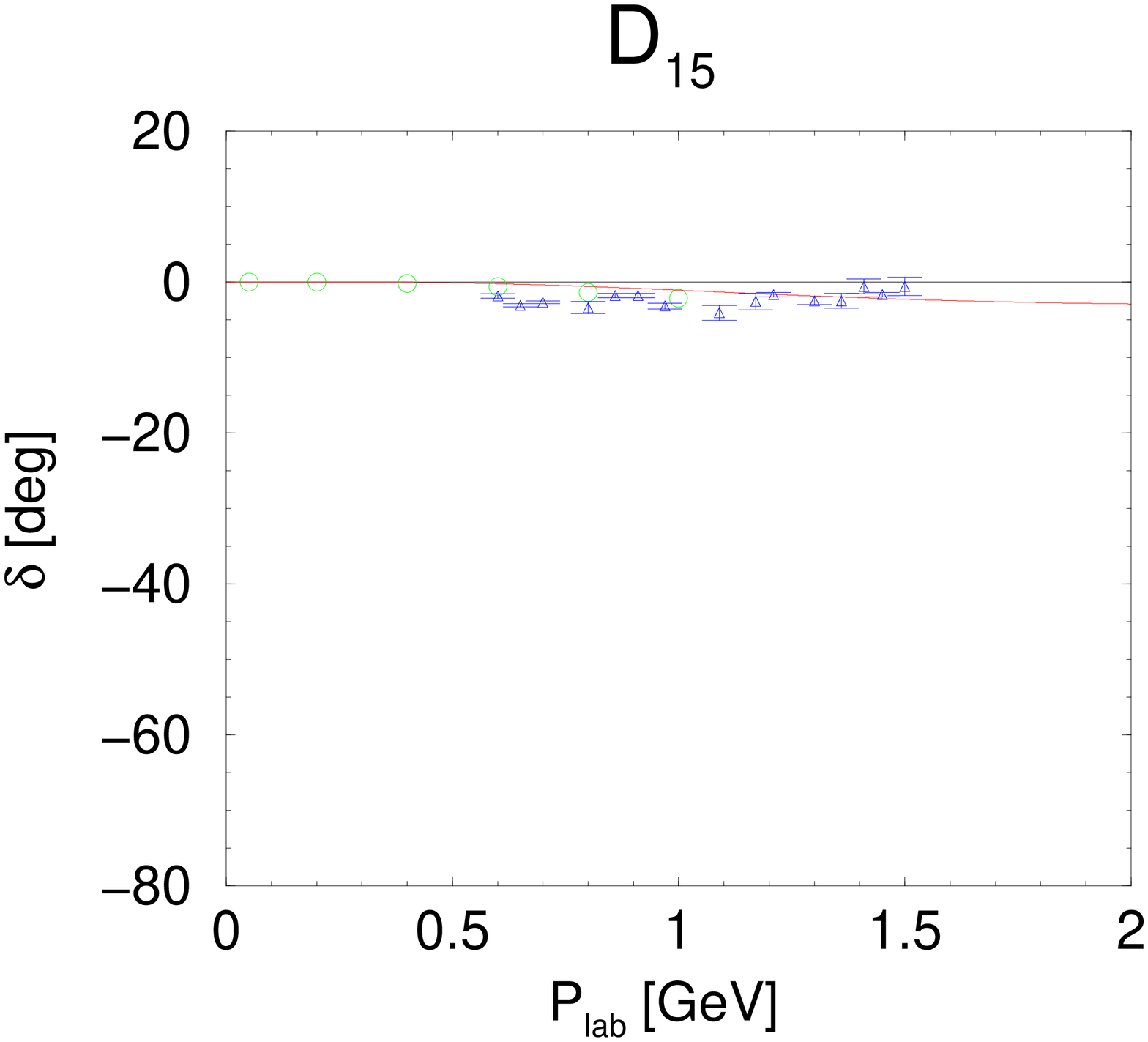,width=1.5in,angle=-90}
\caption{\label{Ieq1}Theoretical I=1 $KN$ phase shifts (red lines).  The experimental phase shifts of
Hashimoto [5] (blue triangles with error bars) and the RGM theoretical phase shifts of Lemaire et al [6]
(green circles and squares, representing two wavefunctions) are shown for comparison.}
\end{figure}

\begin{figure}
\epsfig{figure=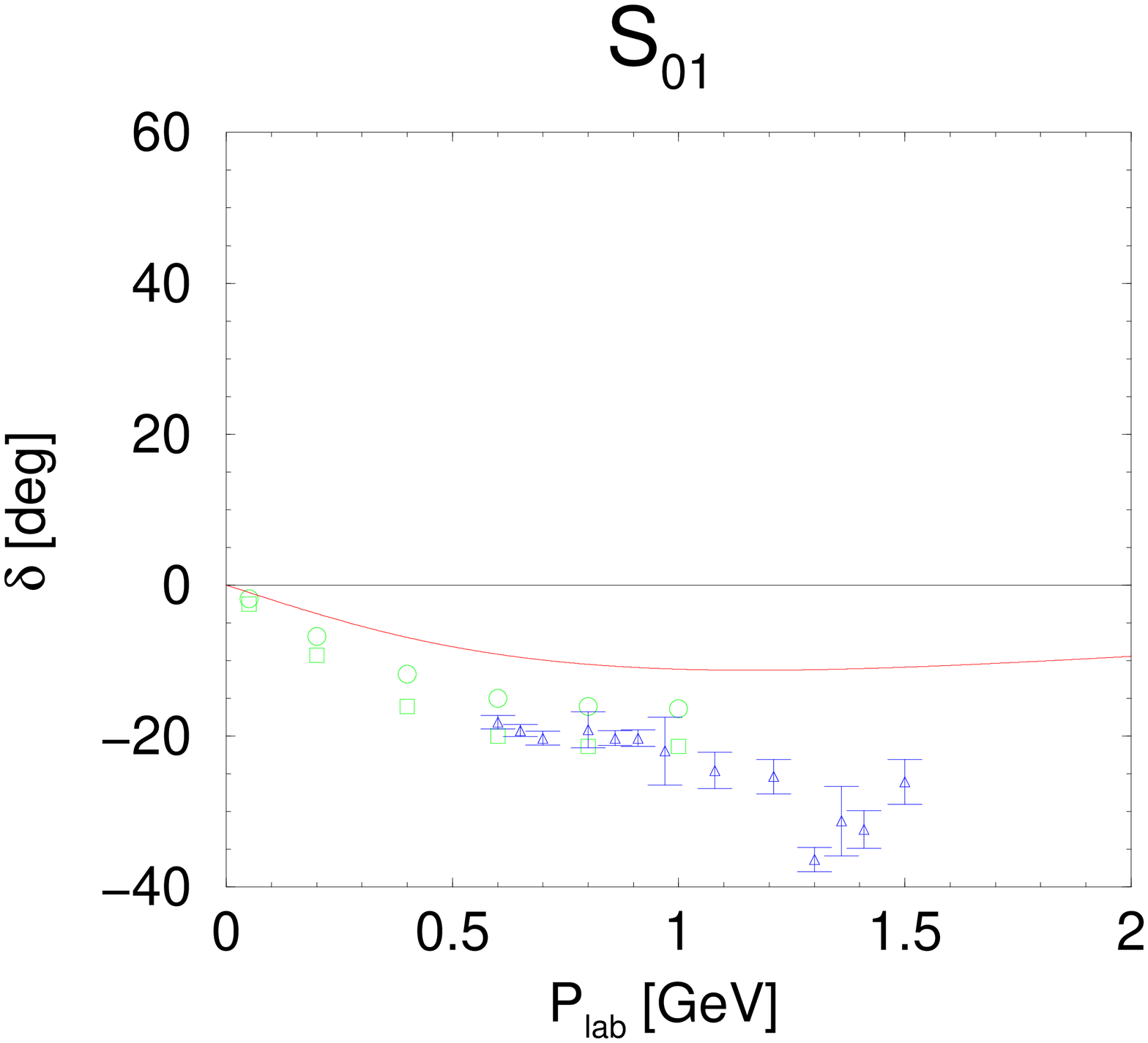,width=1.5in,angle=-90}
\hspace{1cm}
\epsfig{figure=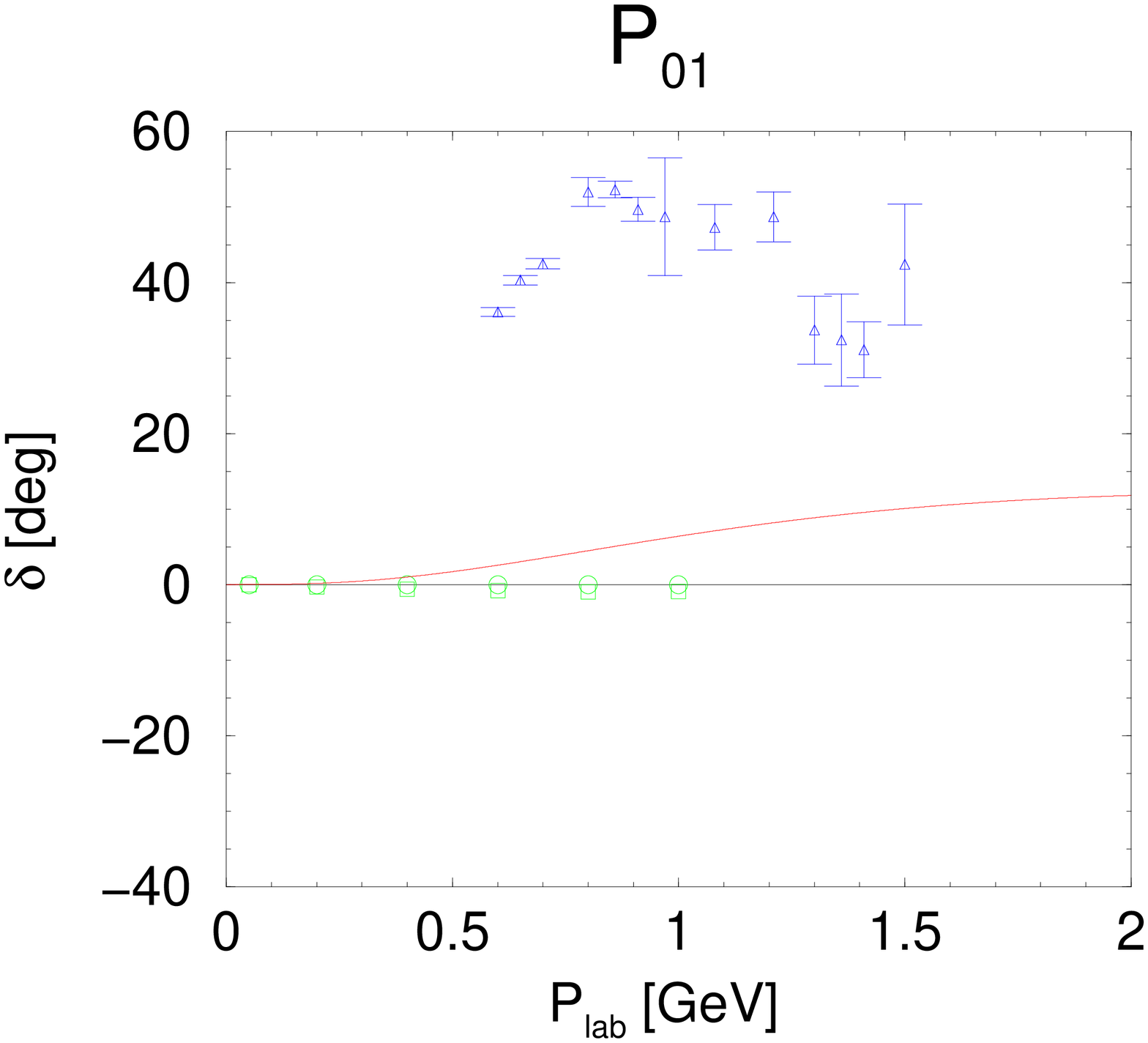,width=1.5in,angle=-90}
\hspace{1cm}
\epsfig{figure=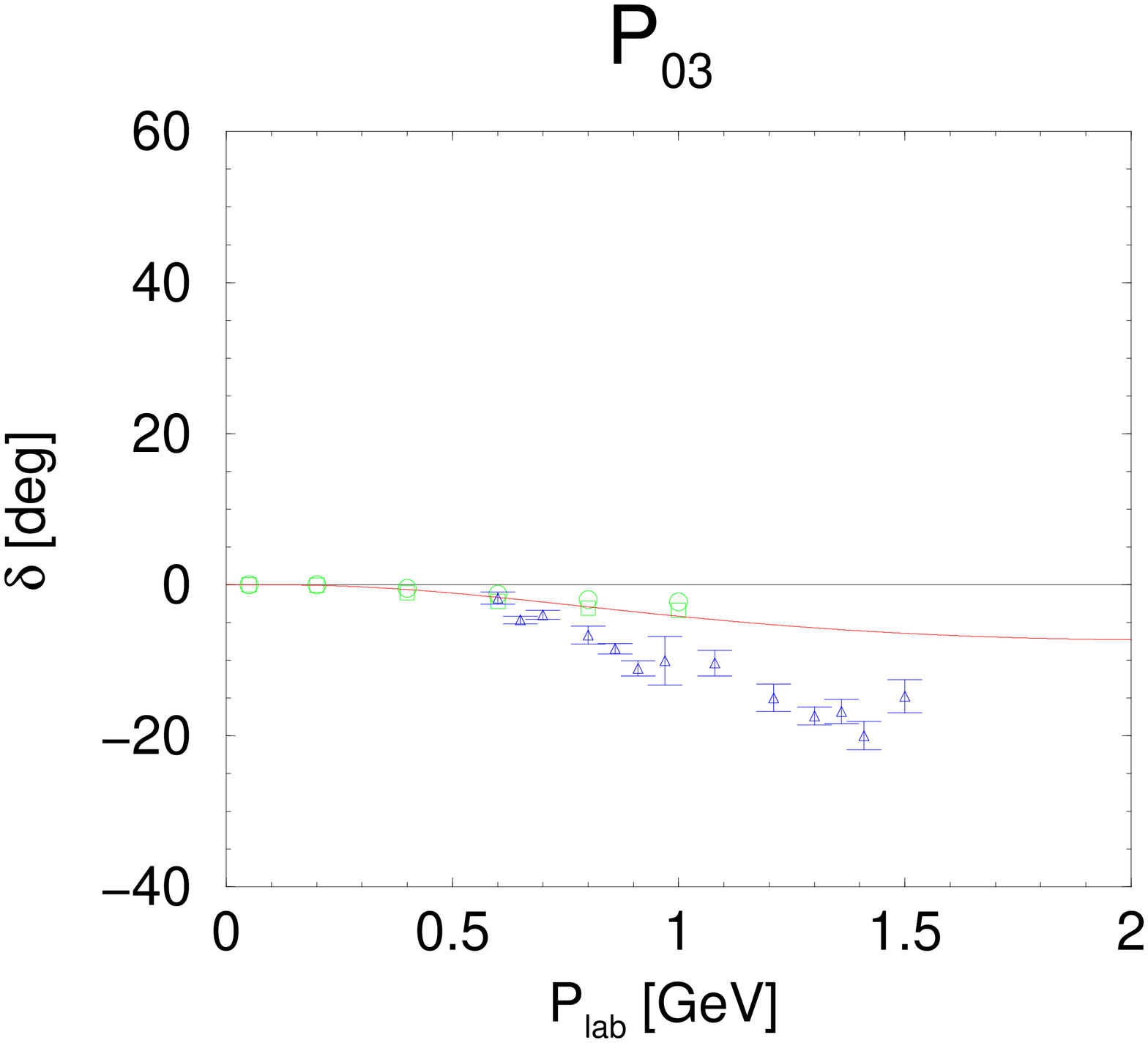,width=1.5in,angle=-90}

\vspace{0.5cm}

\epsfig{figure=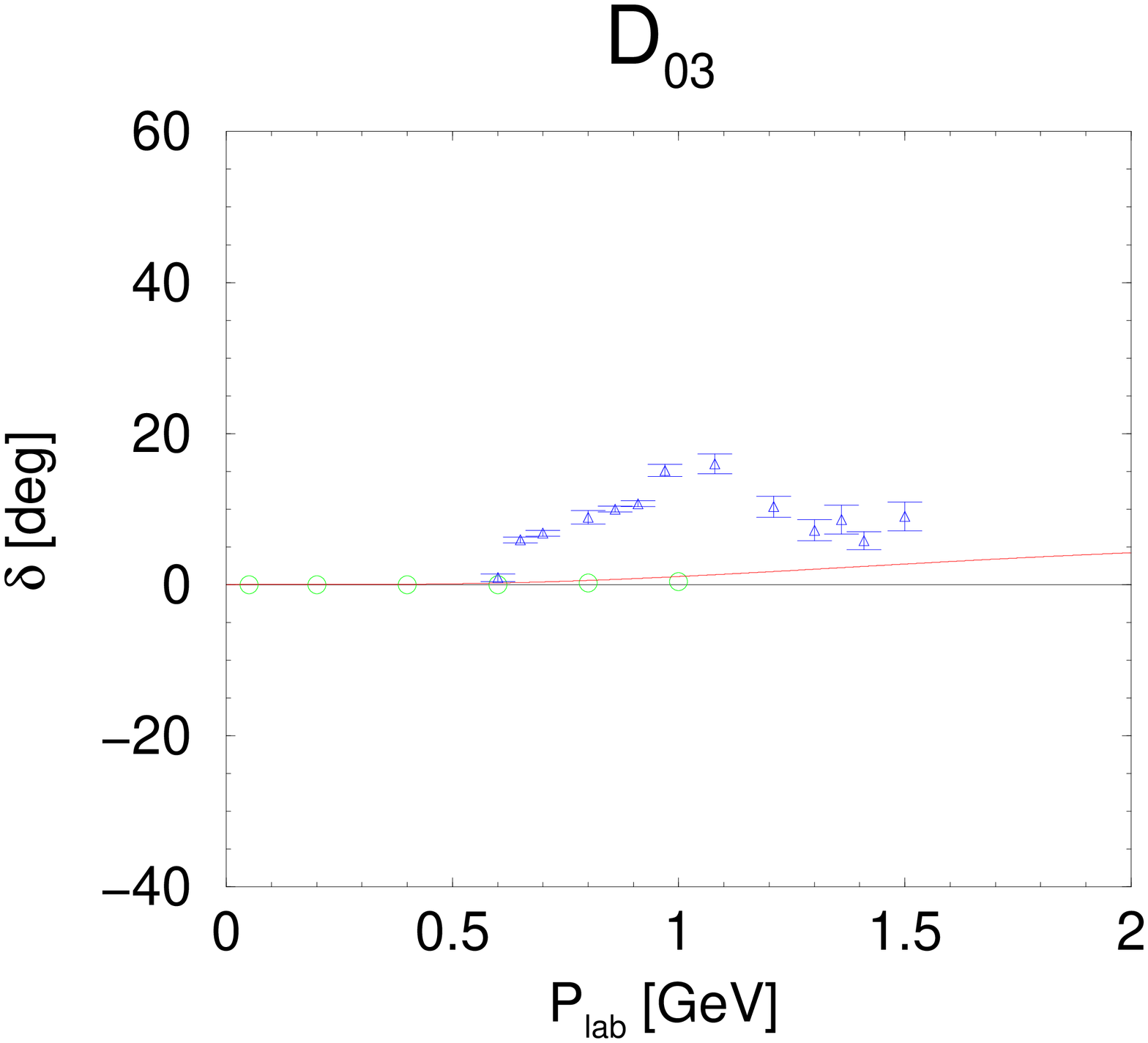,width=1.5in,angle=-90}
\hspace{1cm}
\epsfig{figure=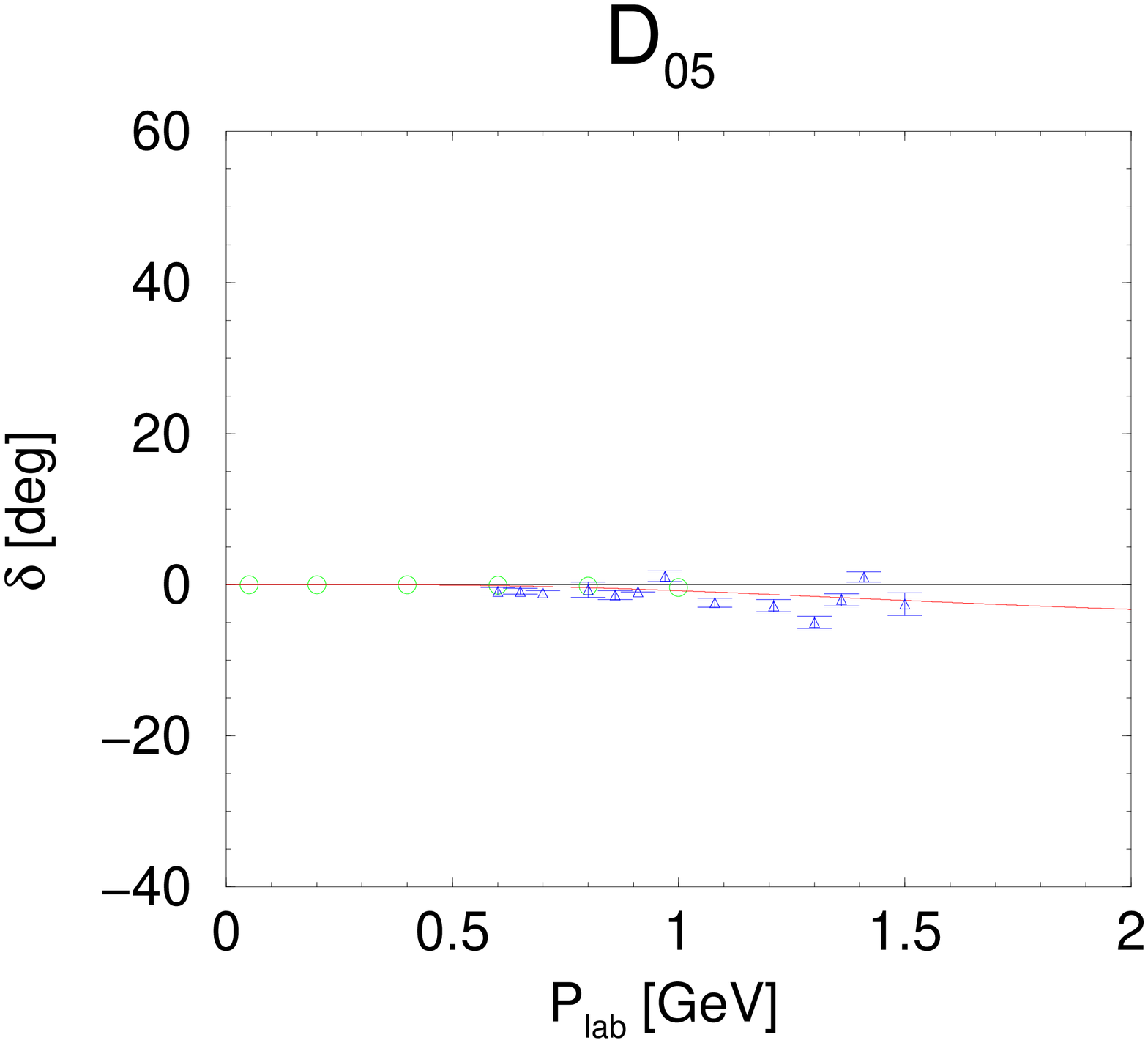,width=1.5in,angle=-90}
\caption{\label{Ieq0}I=0 $KN$ phase shifts, legend as in Figure 1.}
\end{figure}

\begin{figure}
\epsfig{figure=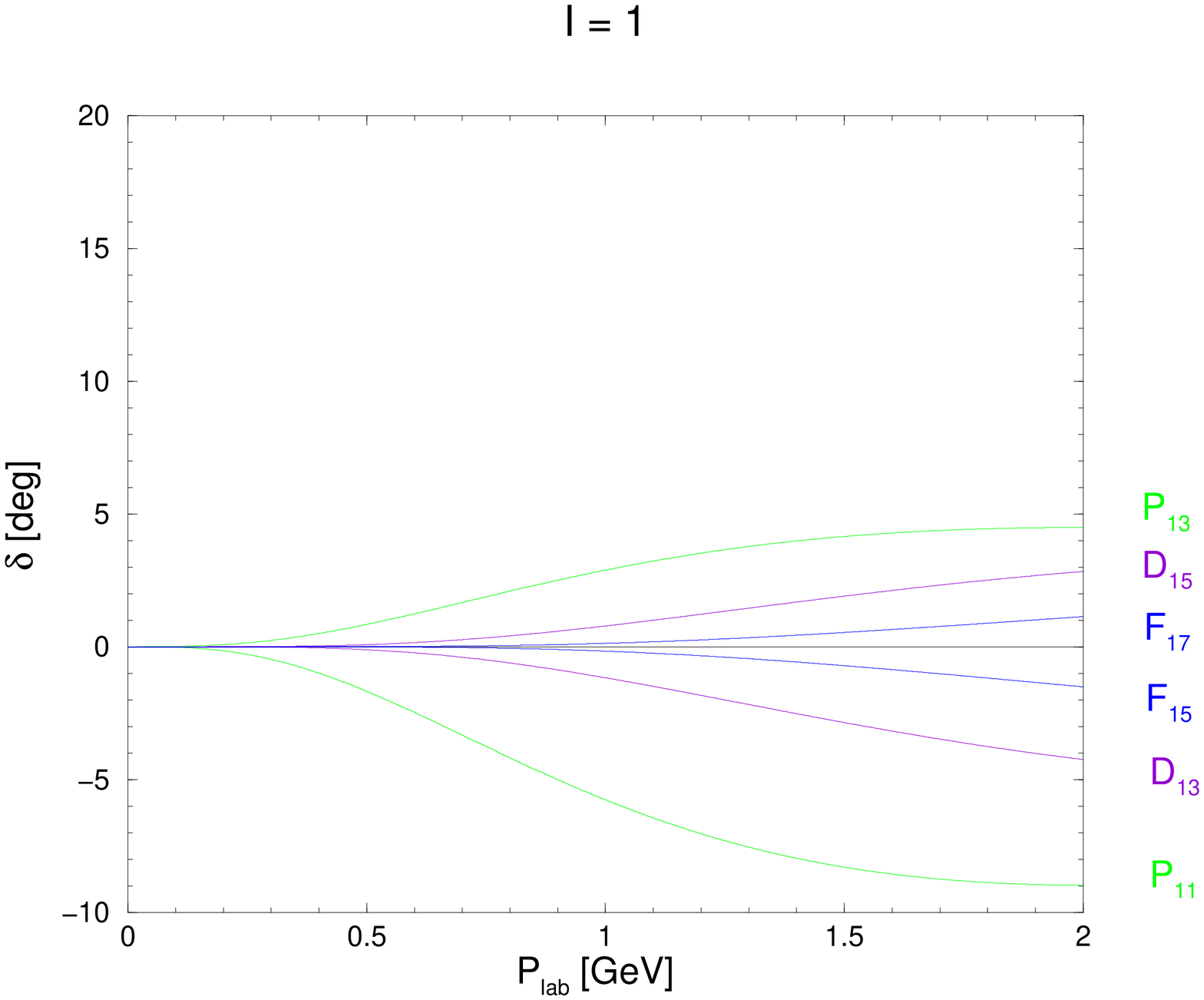,width=2.4in,angle=-90}
\hspace{1cm}
\epsfig{figure=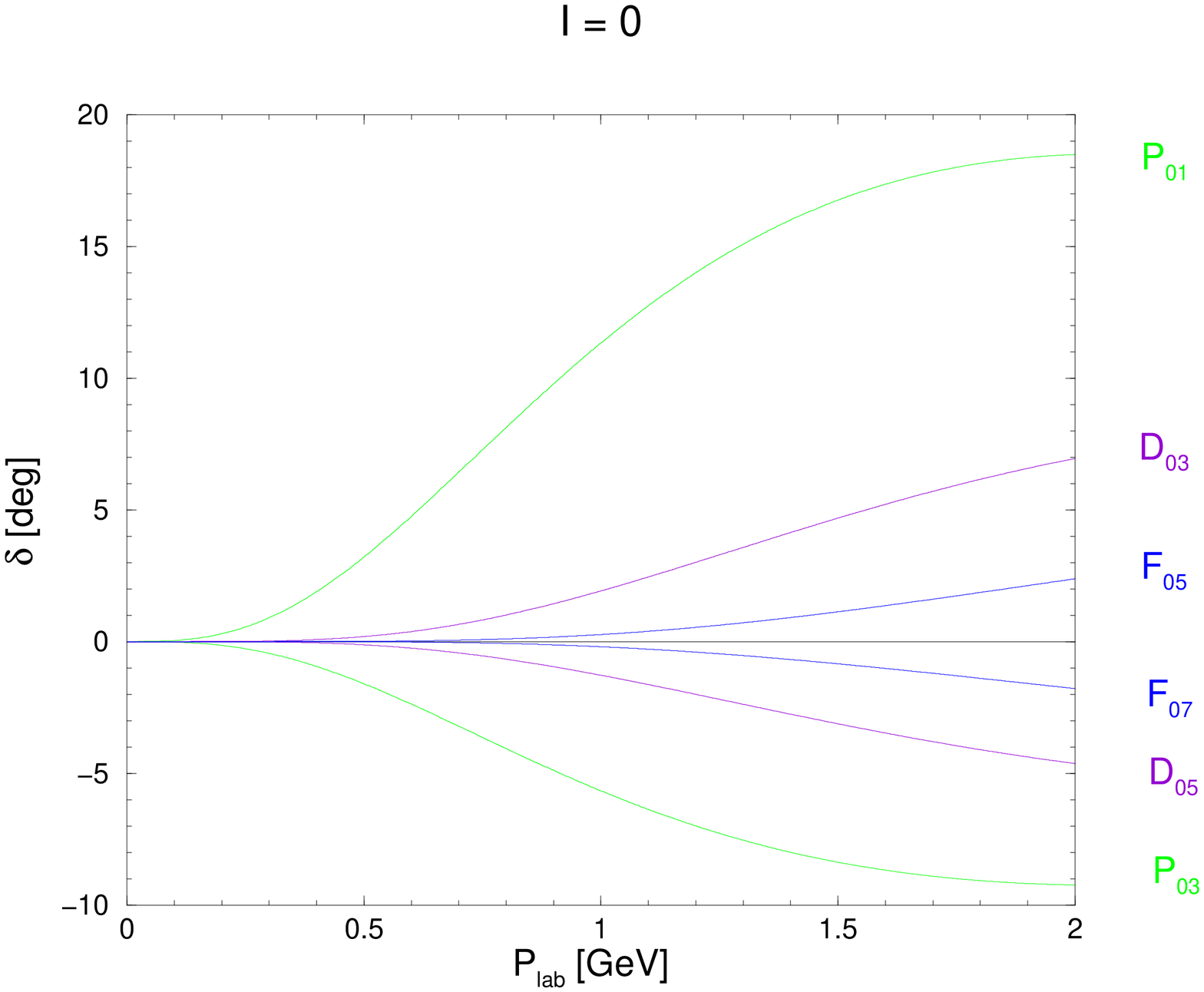,width=2.4in,angle=-90}
\caption{\label{KNogeLS}Theoretical $KN$ OGE spin-orbit phase shifts.}
\end{figure}

\begin{figure}
\epsfig{figure=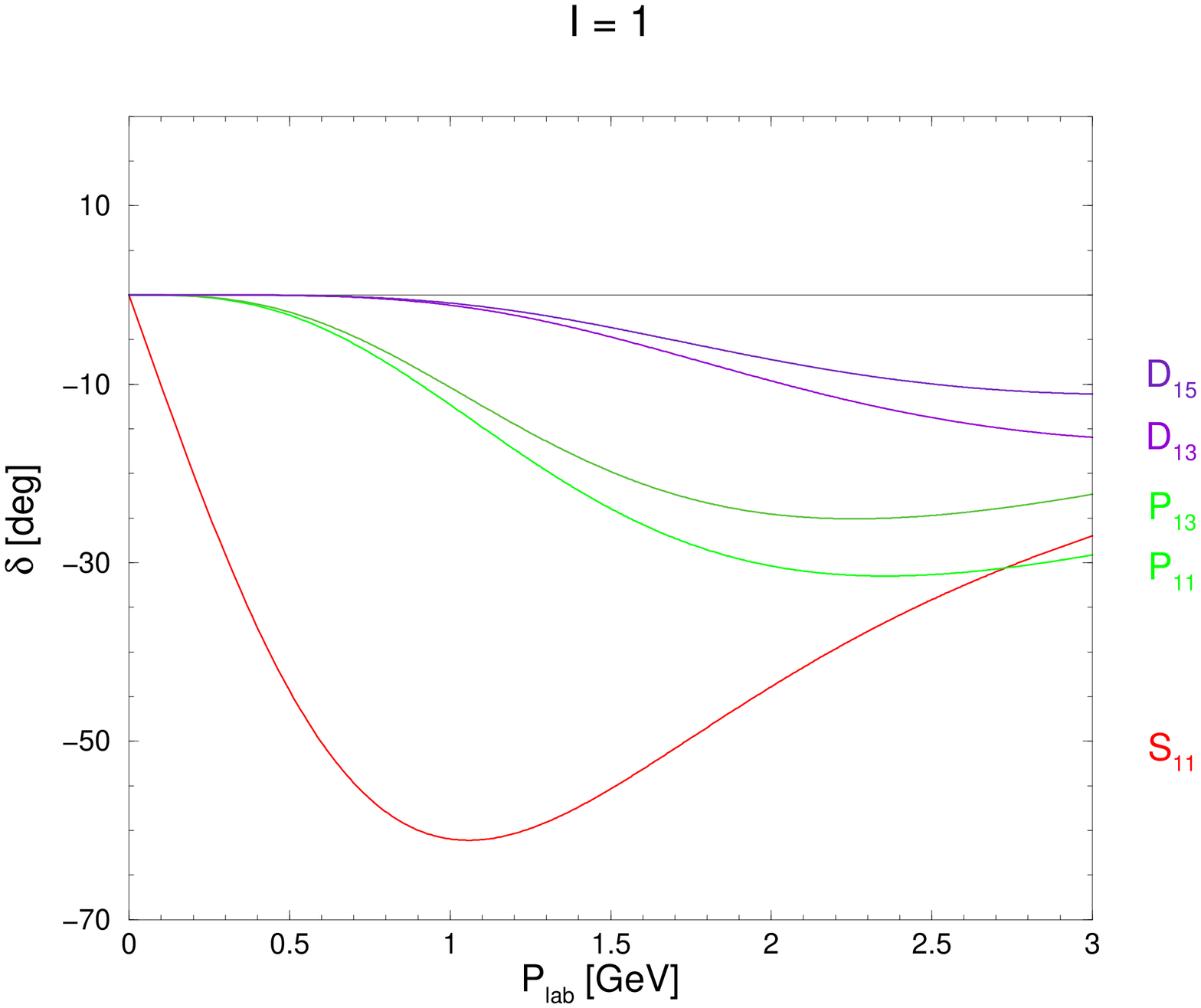,width=2.4in,angle=-90}
\hspace{1cm}
\epsfig{figure=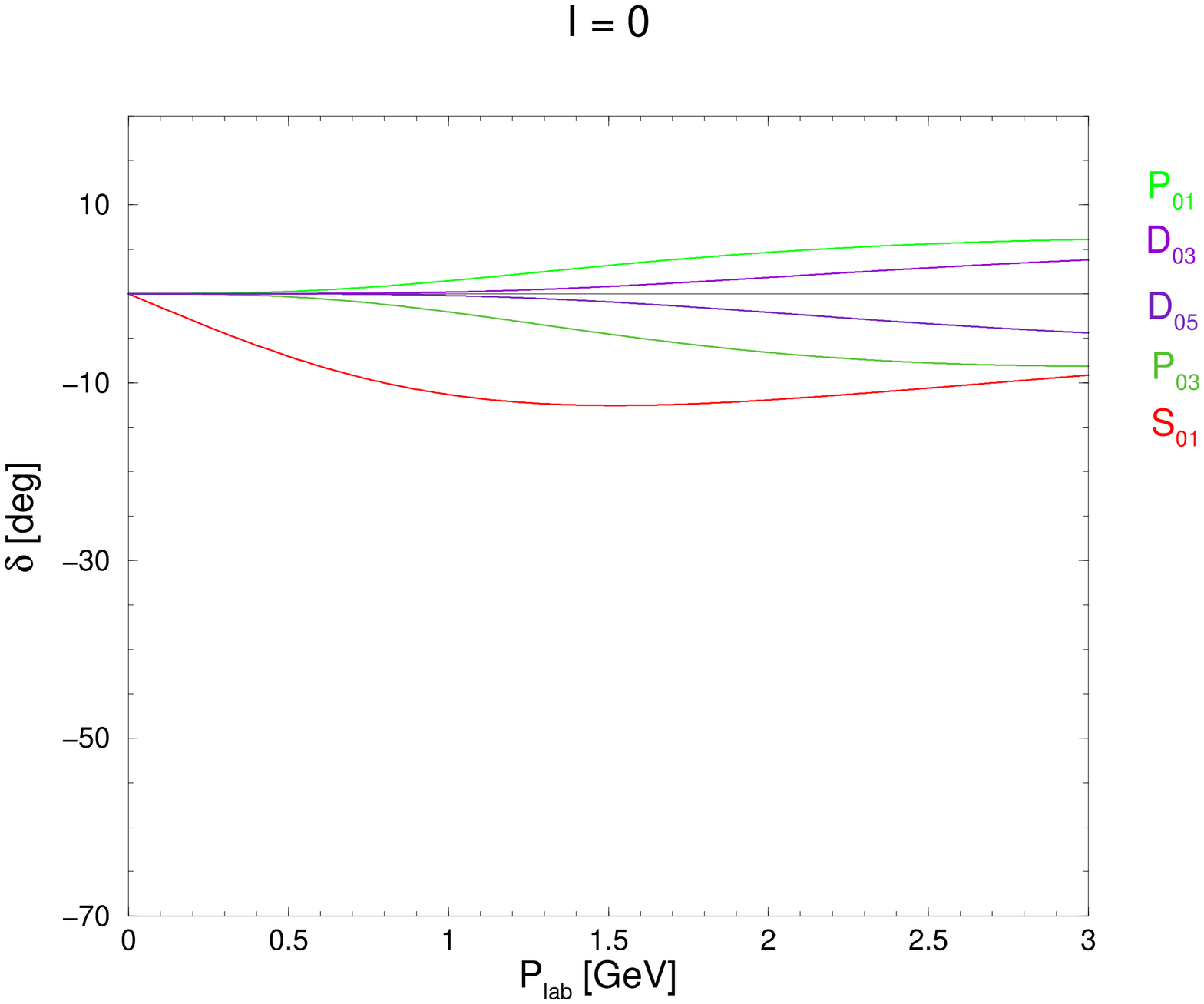,width=2.4in,angle=-90}
\caption{\label{DN}Theoretical $DN$ phase shifts.}
\end{figure}

\end{document}